\documentstyle[aps,preprint]{revtex}

\def\boxit#1{\vbox{\hrule height1pt\hbox{\vrule width1pt\kern10pt
           \vbox{\kern10pt#1\kern10pt}\kern10pt\vrule width1pt
               }\hrule height1pt}}

\begin{document}

\def\power#1{\mbox{$\times10^{#1}\ $}}
\newcommand{\F}{\mbox{$^{19}$F}}
\newcommand{\Ne}{\mbox{$^{19}$Ne}}
\newcommand{\N}{\mbox{$^{15}$N}}
\newcommand{\Li}{\mbox{$^{7}$Li}}
\newcommand{\reac}{$^{15}$N($\alpha,\gamma)^{19}$F}
\newcommand{\reacm}{$^{15}$O($\alpha,\gamma)^{19}$Ne}
\newcommand{\react}{$^{15}$N($^7$Li,t)$^{19}$F}
\newcommand{\degr}{$^\circ$~}
\newcommand{\zaa}{Astron. Astrophys.~}
\newcommand{\zapj}{Astrophys. J.~}
\newcommand{\znp}{Nucl.~Phys.}
\newcommand{\zpl}{Phys.~Lett.~}
\newcommand{\zpr}{Phys.~Rev.~}
\newcommand{\znim}{Nucl.~Inst.~and~Meth.}
\newcommand{\zADNDT}{Atomic Data and Nuclear Data Tables}
\newcommand{\zup}[1]{$\mathrm{^{#1}}$}

\title{Comparison of low--energy resonances in \reac\ and \reacm\
and related uncertainties}

\author{F.~de~Oliveira\footnote{present address: GANIL, B.P. 5027,
F-14021 Caen Cedex, France}
, A.~Coc, P.~Aguer, G.~Bogaert, J.~Kiener,
A.~Lefebvre, V.~Tatischeff, and J.-P.~Thibaud\\
}
\address{CSNSM, IN2P3-CNRS, 91405 Orsay Campus, France}

\author{S.~Fortier, J.M.~Maison, L.~Rosier, G.~Rotbard, and 
J.~Vernotte}
\address{IPN, IN2P3-CNRS et Universit\'e de Paris-Sud, \\
91406 Orsay Cedex, France}

\author{S.~Wilmes, P.~Mohr\footnote{present address:
Institut f\"ur Kernphysik, Technische Universit\"at Wien,
Wiedner Hauptstra{\ss}e 8--10, A--1040 Vienna, Austria}, 
V.~K\"olle, and G.~Staudt}
\address{Physikalisches Institut, Universit\"at T\"ubingen,\\
Auf der Morgenstelle 14, D--72076 T\"ubingen, Germany}

\date{\today}

\maketitle

\begin{abstract}
A disagreement between two determinations of $\Gamma_\alpha$
of the astrophysically relevant level at $E_x$=4.378 MeV in $^{19}$F
has been stated in two recent papers by Wilmes et al. and de~Oliveira et al.
In this work the uncertainties of both papers are discussed in detail, 
and we adopt the value
$\Gamma_\alpha$ = $(1.5^{+1.5}_{-0.8})$\power{-9}~eV for the 4.378 MeV state.
In addition, the validity and the uncertainties of the usual
approximations for mirror nuclei
$\Gamma_\gamma (^{19}{\rm F}) \approx \Gamma_\gamma(^{19}{\rm Ne})$,
$\theta^2_\alpha (^{19}{\rm F}) \approx \theta^2_\alpha (^{19}{\rm Ne})$
are discussed, together with
the resulting uncertainties on the resonance strengths in $^{19}$Ne
and on the \reacm\ rate. 
\end{abstract}

\narrowtext

In a recent publication, Wilmes et al.\cite{Wil95} present
experimental and theoretical results on the \reac\ reaction which is
crucial for fluorine production in AGB stars\cite{Mow96}.
In their experiment, Wilmes et al.\cite{Wil95} used a windowless \N\
gas target and a high purity Ge detector covering the angles
between 60\degr and 120\degr.
They determined for the first time the strength of the
$E_{\alpha,lab}$ = 687~keV resonance ($E_x$=4.556~MeV \F\
level) relative to the strength $\omega\gamma$ =
(97$\pm$20)~$\mu$eV\cite{Mag87} of the
$E_{\alpha,lab}$ = 679~keV resonance ($E_x$=4.550~MeV level).
Their result $\omega\gamma$ = (8$\pm$3) $\mu$eV is in good agreement
with the previous upper limit of 10~$\mu$eV given by Magnus et 
al.\cite{Mag87}.

De~Oliveira et al.\cite{Oli96} have also investigated the $\alpha$ capture
on \N\ and extracted the $\alpha$ widths
$\Gamma_\alpha$ of some levels in \F .
This experiment used a confined \N\ gas target and a 27.3~MeV \Li\ 
beam to study the \react\ transfer reaction. The resulting tritons 
were analyzed by a split-pole magnetic spectrometer and detected in 
the focal plane by a multiwire drift chamber giving position and angle 
informations.
Finite range DWBA analysis was used to extract the
$\Gamma_\alpha$ of levels. Great care was paid to the study of the
influence of the various parameters entering in the analysis.
Furthermore, 
experiments\cite{Oli96,kious} using solid targets 
(\N\ enriched melamine) were carried out.
In all these transfer experiments it was shown
that the reactions were essentially direct.

Some discrepancies between the results presented in these papers
have been stated \cite{Wil95}. In this Brief Report, the authors of both
series of papers, analyze together the reasons for these discrepancies.

We start with the discussion of the $\alpha$ width
$\Gamma_\alpha$ of the 
$E_x$=4.378~MeV level in \F\ which is of astrophysical interest.
The corresponding resonance dominates the reaction rate
at the typical temperatures of thermal 
pulses in AGB stars\cite{Mow96} (T$\approx2.\times10^8$~K).
The experimental value 
$\Gamma_\alpha$ = $(1.5^{+1.5}_{-0.8})$\power{-9}~eV, 
deduced from the transfer 
experiment\cite{Oli96}, is 60 times
lower than the estimate used by Caughlan and Fowler in their last 
compilation\cite{CF88} where they assumed a value equal to 10\% of 
the Wigner limit ($\theta^{2}$ = 0.1).
Wilmes et al.\cite{Wil95}
provided no {\em experimental} information on the
$E_x$=4.378~MeV level.
They assume the identity of the alpha
structure for the $E_x$=4.378~MeV and $E_x$=4.550~MeV levels
and hence the equality of the reduced alpha--widths 
$\theta^2_\alpha$ of both states.
With this assumption, they derive the value
$\Gamma_\alpha$ = 2.4\power{-8}~eV, higher 
by more than one order of magnitude
than the result of de~Oliveira et al.\cite{Oli96}.

The argument of Wilmes et al.\cite{Wil95} is that both levels
belong to the same K$^\pi$=3/2$^+$ band and have the same cluster structure:
$^{12}$C$\otimes^7$Li, quoting Descouvemont and Baye\cite{Des87} who 
also propose $^{11}$B$\otimes^8$Be while Wiescher et al.\cite{Wie80} 
favor $^{14}$N$\otimes^5$He. However, everybody agrees that the 
$^{15}$N$\otimes^4$He component contributes very little to the global 
wave function. 
In these conditions, it appears too simplistic to assume equal alpha
reduced--widths.
Furthermore, the hypothesis of equal reduced widths within the 
K$^\pi$=3/2$^+$ band agrees with the results of Pringle 
and Vermeer\cite{Pri89} only to within a factor of 10.
They measured the
$\Gamma_\gamma/\Gamma$ ratio for various \F\ levels, deduced the
$\Gamma_\alpha$ from the previously known $\omega\gamma$ and
calculated the corresponding reduced widths 
$\Gamma_\alpha/\Gamma_\alpha(s.p.)$.
The values corresponding to four members of the K$^\pi$=3/2$^+$ band:
$E_x(J^\pi)$ =4.55(5/2$^+$), 6.50(11/2$^+$), 6.59(9/2$^+$) and 
10.43(13/2$^+$)~MeV they provide are: $\Gamma_\alpha/\Gamma_\alpha(s.p.)$= 
$(1.1\pm0.2)\times10^{-1}$,
$\ge8.2\times10^{-3}$, $(1.8\pm0.4)\times10^{-2}$  and
$(2.\pm0.5)\times10^{-2}$ respectively. 

The $(^7$Li,t) transfer reaction has been recognized as
a powerful tool to analyze alpha
structures~\cite{Bec78}\cite{Cob76}\cite{Kub72a}\cite{Kub72b} and 
experimental data show that the two levels have not the same
alpha strength. 
This can be seen directly in fig.~4.b of
ref.~\cite{Oli96} where the triton energy spectrum for both levels is
displayed.
(The peaks corresponding to the $E_x$=4.550 and 4.556~MeV levels are 
not resolved but the second is known to be much weaker than the
first\cite{Mag87}\cite{Wil95}.)
Experiments using solid targets \cite{Oli96,kious}
also showed large differences in the $E_x$=4.378 MeV and
$E_x$=4.550 MeV formation through alpha transfer.
The ratio between the 
$E_x$=4.550 and 4.378~MeV reduced widths 
obtained in transfer experiments is 7.5 \cite{Oli96} and therefore within 
the dispersion of values obtained by Pringle and Vermeer \cite{Pri89}
for the same \F\ band.

As result of this discussion we adopt the value of
$\Gamma_\alpha$ = $(1.5^{+1.5}_{-0.8})$\power{-9}~eV, obtained from the
$\alpha$--transfer experiment,
for the alpha width of the level $E_x$ = 4.378 MeV of astrophysical interest.
From the transfer
study summarized in ref.\cite{Oli96} and fully discussed in 
ref.\cite{these},
an uncertainty of factor of two on $\Gamma_\alpha(4.378)$ was estimated.
Together with $\Gamma_\gamma/\Gamma > 0.96$ \cite{Pri89} a resonance
strength is obtained: $\omega \gamma = 6^{+6}_{-3}$ neV.

In contrast to the $E_x$=4.378 MeV level, for the $E_x$=4.550 MeV, 4.683 MeV,
and 5.107 MeV levels a comparison between experimental results of
alpha transfer \cite{Oli96} and alpha capture \cite{Wil95,Wil96,Wil96b}
is possible. The comparison is made in Table~\ref{tab:tab1}, column 5,7 and 10.

In column (3) we list the total widths $\Gamma$ of these states
extracted from the Master Table of Tilley et al.\cite{Til95}.
The value for the 4.550 MeV level is deduced from a lifetime measurement
by Kiss et al.\cite{Kiss82}. Additionally, in the work of Endt \cite{Endt79}
one can find a reduced transition strength for the ground state transition:
$S(\gamma_0)=1.0 \pm 0.2 \, W.u. \;$ i.e. $\Gamma_{\gamma_0}=4.8 \pm 1.0$ meV.
Together with the branching ratio of $4 \pm 2$\% for this 
transition \cite{Wil96b} one obtains $\Gamma_\gamma = 120 \pm 60$ meV
which agrees reasonably well with the result of Kiss et al.\cite{Kiss82}.

In columns (4) and (5) the results of Pringle and Vermeer \cite{Pri89}
are given: They determined the alpha widths $\Gamma_\alpha$ from their
measured branchings $\Gamma_\gamma/\Gamma$  and the previously known
resonance strengths.

In columns (6) and (7) we list the reduced alpha widths $\theta^2_\alpha$
derived from the alpha transfer experiment \cite{Oli96} and the
extracted alpha widths $\Gamma_\alpha$.

Columns (8), (9), and (10) 
show the results of Wilmes et al.\cite{Wil96,Wil96b}.
The resonance strengths $\omega \gamma$ of 14 resonances in \reac\ have 
been measured directly; using the branching ratios $\Gamma_\gamma/\Gamma$
from Ref.~\cite{Pri89}, values for $\Gamma_\alpha$ and $\theta^2_\alpha$
have been deduced. (Note that $\omega \gamma$(4.556) differs slightly
from the value given in Ref.~\cite{Wil95}: 
the previous value was determined from 
an experiment by Magnus et al.\cite{Mag87}, the new value
is the result of the absolute determination.)

For both levels $E_x$=4.550 MeV and $E_x$=4.683 MeV the agreement
between the alpha-transfer \cite{Oli96} and the alpha--capture
results \cite{Wil95,Mag87,Wil96,Wil96b} is good within a factor of 2.
But in the case of the $E_x$=5.107 MeV level the results disagree by
one order of magnitude.
This level is the most weakly produced, has the highest compound
nucleus contribution and the highest excitation energy of the levels
studied by de~Oliveira et al.\cite{Oli96}.
This last point is likely the source of the discrepancy since the FR--DWBA
analysis was performed by de~Oliveira et al. within the approximation that
the relevant levels are bounds. 
However, in the case of the $E_x$=5.107 MeV level,
unbound by $\approx$1.1~MeV this approximation is questionable. On the
contrary, Table~\ref{tab:tab1} shows a reasonable agreement for the low lying
levels when comparison is possible.

In addition, for the 4.378~MeV level,
Wilmes et al.\cite{Wil95} pointed out an apparent contradiction
between the value $\Gamma_\alpha$ deduced from the alpha-transfer experiment
on $^{15}$N
and a value deduced from the ratio $\Gamma_\alpha/\Gamma$ obtained by
Magnus et al.\cite{Mag90} in the mirror nucleus $^{19}$Ne.
This estimate is based on both the assumptions that M1
transition strengths and reduced alpha widths are equal for analog
levels.
The first hypothesis is known to be an useful approximation for
moderately strong M1 transitions where the isovector spin contribution
is expected to dominate.
But for such M1 transitions, unlike in the E1 ones, the
{\em quasi--rule} is that transitions in conjugate nuclei are
expected to be of approximately equal strengths only to within a
factor of $\approx$2.\cite{WW69}.
This {\em quasi--rule} is further broken in most cases when considering
isospin mixing (see for instance, a recent comparison of
transition strengths in $^{15}$O and $^{15}$N by Raman et
al.\cite{Ram94}).
The hypothesis of equality of alpha reduced widths in analog levels
is a quite common practice in the absence of direct measurement.
But it is clearly an approximation whose precision is hard to estimate but
would become more and more questionable when the alpha--structure of the
involved levels is getting weaker.

The alpha strength of the levels under consideration is weak.
One can combine the experimentally available data on $^{19}$F with
the $\Gamma_\alpha/\Gamma$ data from Magnus et al.\cite{Mag90} in \Ne\ to 
test the hypothesis of equal $\theta^2_\alpha$ values in the mirror
nuclei \F\ and \Ne\ assuming the equality of $\Gamma_\gamma$
values for the mirror states (as mentioned above, this assumption
is questionable). The results are listed in Table~\ref{tab:tab2}.
Columns (1) -- (4) give $E_x$, $J^\pi$, and $\Gamma_\gamma$,
column (5) gives the experimental values for $B_\alpha = \Gamma_\alpha/\Gamma$
in $^{19}$Ne \cite{Mag90}. In columns (6) and (7) we calculate 
$\Gamma_\alpha = \Gamma_\gamma \cdot B_\alpha / ( 1 - B_\alpha )$ in $^{19}$Ne
and $\theta^2_\alpha$($^{19}$Ne). The reduced widths
$\theta^2_\alpha$($^{19}$F) are given in column (8); the values are taken
from columns (6) and (10) of Table~\ref{tab:tab1}. One can see that the
disagreement exceeds one order of magnitude! Because of the missing 
experimental information on the resonance strengths in \Ne\, the
approximations 
$\Gamma_\gamma (^{19}{\rm F}) \approx \Gamma_\gamma(^{19}{\rm Ne})$ and/or
$\theta^2_\alpha (^{19}{\rm F}) \approx \theta^2_\alpha (^{19}{\rm Ne})$
have been used in several papers. From the new experimental results
one can estimate the validity of these approximations for the mirror nuclei
\F\ and \Ne : the resulting resonance strengths in \Ne\ are uncertain
by at least a factor of 10.

In conclusion, the resonance strengths in \F\ are well-established
within an uncertainty of a factor of 2.
Hence, the value $\Gamma_\alpha$ = $(1.5^{+1.5}_{-0.8})$\power{-9}~eV is 
adopted
for the 4.378 MeV state in \F\, excluding the value used by Caughlan and 
Fowler in their compilation\cite{CF88}.
However, in the case of \Ne\ the resonance strengths remain very uncertain
because the validity of the usual approximations
$\Gamma_\gamma (^{19}{\rm F}) \approx \Gamma_\gamma(^{19}{\rm Ne})$ and
$\theta^2_\alpha (^{19}{\rm F}) \approx \theta^2_\alpha (^{19}{\rm Ne})$
is questionable.
Hence, it results that the \reacm\ rate,
which relies on $\alpha$-transfer data on the mirror
nucleus \N, is not known to a precision better than one order of magnitude.

\squeezetable

\begin{table}
\caption{Properties of some levels in \F\ corresponding 
	to resonances in \reac }
\begin{tabular}{ccc|cc|cc|ccc}
\hline
E$_x$&  J$^\pi$   
	& $\Gamma$\tablenotemark[1]       
	& $\Gamma_\gamma/\Gamma$\tablenotemark[2] 
	& $\Gamma_\alpha$\tablenotemark[2]
	& $\theta^{2}_{\alpha}$\tablenotemark[3]\tablenotemark[4]
	& $\Gamma_\alpha$\tablenotemark[3]
	& $\omega \gamma$\tablenotemark[5]
	& $\theta^{2}_{\alpha}$\tablenotemark[4] 
	& $\Gamma_{\alpha}$\tablenotemark[5] \\
(MeV)&            & (meV)           &                        
	& (meV)                          & (\power{-2})     
	& (meV)                 & (meV)            & (\power{-2})      &
(meV)\\
\hline
4.378 & (7/2)$^+$  &$>$ 60 & $>$ 0.96 & & 0.56& 1.5 \power{-6} & &  &\\
4.550 & (5/2)$^+$  & 101 $\pm$ 55 & & (32 $\pm$ 7) \power{-3} & 4.2   
& 16 \power{-3}  & (96 $\pm$ 12) \power{-3} & 8.4   & (32 $\pm$ 4) \power{-3}\\
4.556 & (3/2)$^-$  & 38$^{+23}_{-19}$ & & $<$ 3 \power{-3} & &   &
(6.4 $\pm$ 2.5) \power{-3}   & 0.84 & (3.2 $\pm$ 1.3) \power{-3}\\
4.683 & (5/2)$^-$  & 43 $\pm$ 8  & $>$ 0.85 & 2.0 $\pm$ 0.3 & 2.4 & 3.0
& 5.6 $\pm$ 0.6 & 1.5 -- 1.8  & 1.9 -- 2.2\\
5.107 & (5/2)$^+$  & $>$ 22 & 0.97 $\pm$ 0.03 & 4.5 $\pm$ 2.7 & 0.33  & 33
& 9.7 $\pm$ 1.6   & 0.033  & 3.3 $\pm$ 0.6\\
\hline
\end{tabular}
\tablenotetext[1]{from Ref.~\protect\cite{Til95}, 
	$\Gamma(4.550)$ from Ref.~\protect\cite{Kiss82}}
\tablenotetext[2]{from Ref.~\protect\cite{Pri89}}
\tablenotetext[3]{from Ref.~\protect\cite{Oli96}}
\tablenotetext[4]{using $R_N = 5.0$ fm }
\tablenotetext[5]{from Refs.~\protect\cite{Wil96,Wil96b}}
\label{tab:tab1}
\end{table}

\begin{table}
\caption{Properties of some mirror levels in \F\ and \Ne\ corresponding 
	to resonances in \reac\ and \reacm }
\begin{tabular}{ccc|c|cc|cc}
\hline
E$_x$($^{19}$F) & E$_x$($^{19}$Ne) &  J$^\pi$   
	& $\Gamma_{\gamma}$\tablenotemark[1]      
	& B$_{\alpha}$($^{19}$Ne)\tablenotemark[2] 
	& $\Gamma_\alpha$($^{19}$Ne) 
	& $\theta^{2}_{\alpha}$($^{19}$Ne)\tablenotemark[3]
	& $\theta^{2}_{\alpha}$($^{19}$F)\tablenotemark[4] \\
(MeV)& (MeV) & & (meV) & & (meV) & (\power{-2}) & (\power{-2})\\
\hline
4.378 & 4.379 & (7/2)$^+$  & $>$ 60 & 0.044 $\pm$ 0.032 & $>$ 2.8 & $>$ 7.8 & 0.56\\
4.550 & 4.600 & (5/2)$^+$  & 101 $\pm$ 55 & 0.25 $\pm$ 0.04 & 33 $\pm$ 18 & 3.2 & 4 -- 8\\
4.556 & 4.549 & (3/2)$^-$  & 38$^{+23}_{-19}$ & 0.07 $\pm$ 0.03 & 2.9$^{+1.7}_{-1.4}$ & 0.06 & 0.84\\
4.683 & 4.712 & (5/2)$^-$  & 43 $\pm$ 8 & 0.82 $\pm$ 0.15 & 195 $\pm$ 36 & 0.67 & 1.5 -- 2.4\\
5.107 & 5.092 & (5/2)$^+$  & $>$ 22 & 0.90 $\pm$ 0.09 & $>$ 200 & $>$ 0.19 & 0.033 -- 0.33\\
\hline
\end{tabular}
\tablenotetext[1]{assuming 
	$\Gamma_{\gamma}$($^{19}$Ne) = $\Gamma_{\gamma}$($^{19}$F) 
	= $\Gamma$($^{19}$F)
	because $\Gamma_\gamma/\Gamma$($^{19}$F) $\approx 1$ 
	(Ref.~\protect\cite{Pri89})}
\tablenotetext[2]{from Ref.~\protect\cite{Mag90}}
\tablenotetext[3]{using $R_N = 5.0$ fm }
\tablenotetext[4]{from Table~\protect\ref{tab:tab1}, columns (6) and (10)}
\label{tab:tab2}
\end{table}

\end{document}